\documentclass[a4paper,amsmath,amssymb]{jpconf}
\usepackage{graphicx}
\usepackage{amsmath, amsthm, amssymb}
\begin{document}
\title{Deterministic flows of order-parameters in stochastic processes of quantum Monte Carlo method}

\author{Jun-ichi Inoue}

\address{Complex Systems Engineering, Graduate School of Information Science and Technology, Hokkaido University, 
N14-W9, Kita-ku, Sapporo 060-0814, Japan
}

\ead{j$\underline{\,\,\,}$inoue@complex.eng.hokudai.ac.jp}

\begin{abstract}
In terms of the stochastic process of quantum-mechanical version of 
Markov chain Monte Carlo method (the MCMC), we analytically derive  macroscopically 
deterministic flow equations of order parameters such as spontaneous magnetization 
in infinite-range ($d(=\infty)$-dimensional) quantum spin systems. 
By means of the Trotter decomposition, 
we consider the transition probability of Glauber-type dynamics of microscopic states  for the corresponding 
$(d+1)$-dimensional classical system. 
Under the static approximation,  differential equations with respect to macroscopic order 
parameters are explicitly obtained from the master equation that describes the microscopic-law. 
In the steady state, we show that  the equations are 
identical to the saddle point equations for the equilibrium state of the same system. 
The equation for the dynamical Ising model is recovered in the classical limit. 
We also check the validity of the static approximation by making use of computer simulations 
for finite size systems  
and discuss several possible extensions of our approach to disordered spin systems 
for statistical-mechanical informatics. 
Especially, we shall use our procedure to evaluate the
decoding process of Bayesian image restoration. 
With the assistance of the concept of dynamical replica theory (the DRT), 
we derive the zero-temperature flow equation of image restoration measure 
showing some `non-monotonic'  behaviour in its time evolution. 
\end{abstract}
\section{Introduction}
\label{sec:Introduction}
In Bayesian statistics and statistical inference within the framework, to evaluate the expectation 
of microscopic labels (such as `pixels' or `bits') 
over the posterior distribution 
of these microscopic states or to 
calculate the marginal probability is very important procedure 
to solve various problems in the research field of 
`massive' probabilistic information processing \cite{Bishop}. 
For most of the cases, it is very hard for us to carry out the task 
except for several limited cases of solvable probabilistic models. 
Recently, to overcome the difficulty, deterministic approaches based on 
mean-field approximation 
including the belief-propagation succeeded in dealing 
with the problems efficiently \cite{Opper,Jordan}. 

On the other hand, in a lot of research fields, the Markov chain Monte Carlo method (the MCMC) 
has been widely used to calculate these expectations or to make marginal distributions  
by sampling the important states that contribute effectively to the macroscopic quantities \cite{Binder}. 
Generally speaking, the MCMC takes a long time to wait  until the Markovian 
stochastic process 
starts to generate the microscopic states from well-approximated posterior distribution.  
Especially, for some classes of probabilistic models 
which are categorized in the so-called {\it spin glasses} \cite{Mezard,BinderYoung,Random},  the time 
consuming is sometimes very serious to make attempt to proceed 
the desired information processing within a realistic time. 
However, even for such cases, various improvements based on several 
important concepts have been proposed and succeeded in carrying out the numerical calculations \cite{SW,Hukushima}. 
From the view point of statistical physics, transitions between microscopic states are controlled by a specific 
hyper-parameter, namely, `temperature' of the system and 
by cooling the temperature slowly enough during the Markovian 
stochastic process, one can get the lowest energy states efficiently. 
This type of optimization tool based on `thermal fluctuation' is referred to as 
{\it simulated annealing} \cite{Kirkpatrick,Geman}. 

Recently, the simulated annealing has been extended to the quantum-mechanical version. 
This new type of simulated annealing called as {\it quantum annealing} \cite{Kadowaki,Farhi,Morita,SuzukiOkada,Santoro}
is based on the adiabatic theorem of the quantum system that evolves 
according to Schr$\ddot{\rm o}$dinger equations. 
To use the quantum annealing, or more generally, to utilize the quantum fluctuation 
for combinatorial optimization problems or massive information 
processing dealing with a huge number of particles such as image restoration or error-correcting codes,  
the approach by solving the Schr$\ddot{\rm o}$dinger equations is apparently limited and 
we should look for different ways to simulate the quantum systems. 

As the most effective and efficient way, the quantum Monte Carlo method \cite{Suzuki} was established. 
The method is based on the following 
Suzuki-Trotter decomposition for non-commutative two operators ${\cal A}$ and ${\cal B}$ : 
\begin{eqnarray}
{\rm tr}\, {\exp}
\left(
{\cal A}+{\cal B}
\right) & = & 
\lim_{M \to \infty} 
{\rm tr}
\left(
{\exp}
\left(
\frac{{\cal A}}{M}
\right)
{\exp}
\left(
\frac{{\cal B}}{M}
\right)
\right)^{M}
\label{eq:ST}
\end{eqnarray}
Namely, 
the $d$-dimensional quantum 
system is mapped to the corresponding $(d+1)$-dimensional classical spin systems.  
This approach is very powerful and a lot of 
researches have succeeded in exploring  the quantum phases in strongly correlated quantum systems. 
However, when we simulate the quantum system at zero temperature in which quantum effect is essential, 
we encounter some technical difficulties although several sophisticated algorithms were proposed \cite{Oliveira}. 
From the view point of information science, 
it is very informative for us to evaluate  
the process of information processing and 
if one seeks to utilize the quantum fluctuation to solve the problems, 
we should use  the `zero-temperature dynamics'. 
For this purpose, it seems that we need some tractable `bench mark tests'  to 
investigate the `dynamical process' of information processing such as quantum annealing 
by using quantum fluctuation at zero temperature.  

In classical system, 
Coolen and Ruijgrok \cite{Coolen1988} proposed 
a way to derive the differential 
equations with respect to 
order-parameters of the system 
from the microscopic master equations 
for the so-called Hopfield model \cite{Hopfield} as an associative memory in which a finite number of patterns are embedded. 
The procedure was extended by Coolen and Sherrington \cite{Coolen1994}, Coolen, Laughton and Sherrington \cite{Coolen1996}   
to more complicated spin systems categorized in the infinite-range (or mean-field) models 
including the Sherrington-Kirkpatrick spin glasses \cite{SK}. 
The so-called dynamical replica theory (DRT) was now 
established 
as a strong approach to investigate the dynamics in the classical disordered spin systems . 
On the considering the matter, 
it seems to be very important for us to extend their approach to 
quantum systems evolving stochastically according to 
the quantum Monte Carlo dynamics. 
  
In this paper, in terms of the stochastic process of quantum-mechanical version of 
Markov chain Monte Carlo method, we analytically derive macroscopically 
deterministic flow equations of order parameters such as spontaneous magnetization 
in infinite-range ($d(=\infty)$-dimensional) quantum spin systems. 
By means of the Trotter decomposition, 
we consider the transition probability of Glauber-type dynamics of microscopic states  for the corresponding 
$(d+1)$-dimensional classical system. 
Under the static approximation,  differential equations with respect to macroscopic order 
parameters are explicitly obtained from the master equation that describes the microscopic-law. 
In the steady state, we show that  the equations are 
identical to the saddle point equations (equations of states) for the equilibrium state of the same system. 
We easily find that the equation for the dynamical Ising model 
is recovered in the classical limit. 
We also check the validity of the static approximation by computer simulations 
for finite size systems  and discuss several possible extensions of our approach to disordered spin systems 
for statistical-mechanical informatics \cite{Nishi}. 
Especially, we shall use our procedure to evaluate the
decoding process of Bayesian image restoration \cite{TH,NW,Tanaka,Inoue,Inoue2}. 
With the assistance of the concept of dynamical replica theory (the DRT) \cite{Coolen1994,Coolen1996}, 
we derive the zero-temperature flow equation of image restoration measure, 
namely, overlap function 
between a given original image and the degraded one,  
showing some `non-monotonic'  behaviour in its time evolution. 

This paper is organized as follows. 
In Section 2, we explain our formulation 
to describe the deterministic flow equation from 
master equation for a simplest quantum spin system. 
In Section 3, we apply our approach to evaluate the
decoding process of Bayesian image restoration. 
 Some `non-monotonic'  behaviour in 
 time evolution of image restoration measure is observed 
 by the analysis of flow equations. 
 The last section contains some remarks. 
\section{The model system and formulation}
In this section, we derive the differential equations  
with respect to several order parameters  for 
a simplest quantum spin system, namely, 
a class of the infinite range transverse Ising model \cite{Bikas1996,Sachdev} described by 
the following Hamiltonian:  
\begin{eqnarray}
H & = & 
-\frac{1}{N}
\sum_{i,j=1}^{N}
J_{ij}
\sigma_{i}^{z}
\sigma_{j}^{z}- 
h \sum_{i=1}^{N}\tau_{i}\sigma_{i}^{z} 
 - \Gamma \sum_{i=1}^{N}
\sigma_{i}^{x} 
\label{eq:Hamiltonian}.
\end{eqnarray}
where $\sigma_{i}^{z}$ and $\sigma_{i}^{x}$ denote the Pauli matrices given by 
\begin{eqnarray*}
\sigma_{i}^{z} & = & 
\left(
\begin{array}{cc}
1 & 0 \\
0 & -1
\end{array}
\right), \,\,\,\,
\sigma_{i}^{x} = 
\left(
\begin{array}{cc}
0 & 1 \\
1 & 0
\end{array}
\right).
\end{eqnarray*}
It should be noticed that 
from the view point of Bayesian statistics, 
the above Hamiltonian corresponds to 
the logarithm of the posterior distribution. 
For various choices of 
parameters $\{J_{ij}\}, h$ and $\{\tau\}$, 
one can model the problem of information processing appropriately.   
For instance,  for the choice of $J_{ij}=J, h \neq 0$, 
the above Hamiltonian describes 
image restoration \cite{NW,Inoue,Inoue2} from a given set of 
degraded images 
$\mbox{\boldmath $\tau$}=(\tau_{1},\cdots,\tau_{N})$ 
by means of Markov random fields.   
For $J_{ij} \neq 0, h =0$, 
it corresponds to the logarithm of the posterior of 
the Sourlas codes \cite{Sourlas,Inoue2009} sending the parity check of 
two-body interactions  
$\xi_{i}\xi_{j}\,\,\,\forall (i,j)$ through the binary symmetric channel (BSC). 
For the Sherrington-Kirkpatrick model \cite{SK}, we may choose 
$J_{ij}$ obeying the Gaussian with $J_{0}$ mean and $\tilde{J}^{2}$ variance.   
Especially,  
for the choice of the so-called Hebb rule $J_{ij}=\sum_{\mu=1}^{\alpha N}
\xi_{i}^{\mu}\xi_{j}^{\mu}, h=0$, 
it becomes energy function of the Hopfield model \cite{Hopfield} in which extensive 
number of patterns (loading rate $\alpha$) are embedded. 

In this section, we focus on the simplest case of $J_{ij}=J>0, h=0$ 
ferromagnetic transverse Ising model \cite{Suzuki1966,Elliot}. 
Let us start our argument from the effective Hamiltonian 
which is decomposed by the Suzuki-Trotter formula (\ref{eq:ST}). 
\begin{eqnarray}
\beta H & = & 
-\sum_{k=1}^{M}
\sum_{i=1}^{N} \beta \phi_{i}(\mbox{\boldmath $\sigma$}_{k} : \sigma_{i}(k+1)) \sigma_{i}(k) =  
-\frac{\beta J}{MN}
\sum_{k,ij}
\sigma_{i}(k)
\sigma_{j}(k)
-B 
\sum_{k,i}\sigma_{i}(k)\sigma_{i}(k+1)
\label{eq:effective}
\end{eqnarray}
where $k$ means the Trotter index and 
$M$ denotes the number of the Trotter slices. 
We also defined the parameter $B$ as 
$B \equiv  (1/2) \log \coth (\beta \Gamma/M)$ 
and a microscopic spin state on the $k$-th Trotter slice by 
$\mbox{\boldmath $\sigma$}_{k}  \equiv (\sigma_{1}(k), 
\sigma_{2}(k), \cdots, \sigma_{N}(k)),\, 
 \sigma_{i}(k) \in \{+1,-1 \}$. 
\subsection{The Glauber dynamics and its transition probability}
In the expression of the effective Hamiltonian (\ref{eq:effective}), 
$\beta \phi_{i}(\mbox{\boldmath $\sigma$}_{k},\sigma_{i} (k+1))$ is 
a local field on the cite $i$ in the $k$-th Trotter slice, 
which is explicitly given by 
\begin{eqnarray*}
\beta \phi_{i}(\mbox{\boldmath $\sigma$}_{k} : \sigma_{i} (k+1)) & = & 
\frac{\beta J}{MN}
\sum_{j}\sigma_{j}(k) + 
B \sigma_{i}(k+1). 
\end{eqnarray*}
It should be noticed that 
the effective Hamiltonian (\ref{eq:effective}) is 
a classical system under the Trotter decomposition. 
Therefore,  
the quantum Monte Carlo dynamics should be described by the Glauber-type 
stochastic update rule 
whose transition probability is explicitly written by  
\begin{eqnarray*}
w_{i}(\mbox{\boldmath $\sigma$}_{k}) & = & 
\frac{1}{2}
\left[
1-\sigma_{i}(k) \tanh (\beta \phi_{i}(\mbox{\boldmath $\sigma$}_{k} : \sigma (k+1)))
\right]. 
\end{eqnarray*}
Namely, this means that 
\begin{eqnarray}
\sigma_{i} (k) & = & 
\left\{
\begin{array}{ccc}
+1 & \mbox{with prob.} & 
\frac{{\exp} [\beta \phi_{i}(\mbox{\boldmath $\sigma$}_{k} : \sigma_{i} (k+1))]}
{{\exp}[\beta \phi_{i}(\mbox{\boldmath $\sigma$}_{k} : \sigma_{i} (k+1))]
+{\exp}[-\beta \phi_{i}(\mbox{\boldmath $\sigma$}_{k} : \sigma_{i} (k+1))]} \\
-1 & \mbox{with prob.} & 
\frac{{\exp} [-\beta \phi_{i}(\mbox{\boldmath $\sigma$}_{k} : \sigma_{i} (k+1))]}
{{\exp}[\beta \phi_{i}(\mbox{\boldmath $\sigma$}_{k} : \sigma_{i} (k+1))]
+{\exp}[-\beta \phi_{i}(\mbox{\boldmath $\sigma$}_{k} : \sigma_{i} (k+1))]} 
\end{array}
\right.
\end{eqnarray}
Obviously, in the classical limit, namely, 
when $B$ goes to infinity as $\Gamma \to 0$  and 
$\phi_{i}(\mbox{\boldmath $\sigma$}_{k} : \sigma_{i}(k+1)) \sim B\sigma_{i}(k+1)$,  
$\sigma_{i}(k)$ is identical to 
$\sigma_{i}(k+1)$ with 
probability $1$. 
Thus, for small $\Gamma$, we have $\mbox{\boldmath $\sigma$}_{k}=
\mbox{\boldmath $\sigma$}_{k+1}$ with a relatively high probability. 
In this sense, 
we can regard the term $B\sigma_{i}(k+1)$ as 
an `external field' from the nearest Trotter slice $k+1$. 
Here we use the periodic boundary condition for the Trotter direction, 
that is, $\mbox{\boldmath $\sigma$}_{1}=
\mbox{\boldmath $\sigma$}_{M+1}$. 
\subsection{The master equation}
Then, the master equation 
for the probability of the microscopic states  
on the $k$-th Trotter slice $p_{t} (\mbox{\boldmath $\sigma$}_{k})$  
is written by 
\begin{eqnarray}
\frac{dp_{t}(\mbox{\boldmath $\sigma$}_{k})}
{dt} & = & 
\sum_{i=1}^{N}
\left[
p_{t}(F_{i} ^{(k)}(\mbox{\boldmath $\sigma$}_{k}) )
w_{i}(F_{i}^{(k)}(\mbox{\boldmath $\sigma$}))-
p_{t}(\mbox{\boldmath $\sigma$}_{k})
w_{i}(\mbox{\boldmath $\sigma$}_{k})
\right]
\label{eq:master}
\end{eqnarray}
where $p_{t}(\mbox{\boldmath $\sigma$}_{k})$ denotes a 
probability that the system in the $k$-th Trotter slice 
is in a microscopic state $\mbox{\boldmath $\sigma$}_{k}$ at time $t$. 
We also defined an operator $F_{i}^{(k)}$ 
to flip a single spin on the cite $i$ in the $k$-th 
Trotter slice as $\sigma_{i} (k) \to -\sigma_{i} (k)$. 
\subsection{From master equation to deterministic flow of order parameter}
We next introduce the spontaneous magnetization 
in the $k$-th Trotter slice 
$m_{k} = N^{-1} \sum_{i} \sigma_{i}(k)$ 
as a relevant 
order parameter. 
The probability that the system is described by the 
magnetization $m_{k}$ at time $t$ is given 
in terms of the probability $p_{t}(\mbox{\boldmath $\sigma$}_{k})$ for a given 
realization of the microscopic state as $P_{t}(m_{k})  =  
\sum_{\mbox{\boldmath $\sigma$}_{k}} 
p_{t}(\mbox{\boldmath $\sigma$}_{k}) 
\delta (m_{k}-m_{k}(\mbox{\boldmath $\sigma$}_{k}))$. 
Taking the derivative of this equation 
with respect to $t$ and substituting (\ref{eq:master}) into the result, we obtain 
\begin{eqnarray}
\frac{dP_{t}(m_{k})}
{dt} & = & 
\frac{\partial}{\partial m_{k}}
\sum_{\mbox{\boldmath $\sigma$}_{k}}
p_{t}(\mbox{\boldmath $\sigma$}_{k})
\sum_{i}
\frac{\sigma_{i}(k)}{N}
\left[
1-\sigma_{i}(k)
\tanh (\beta \phi_{i}(\mbox{\boldmath $\sigma$}_{k} : \sigma_{i}(k+1)))
\right]
\delta (m_{k}-m_{k}(\mbox{\boldmath $\sigma$}_{k})) \nonumber \\
\mbox{} & = & 
\frac{\partial}{\partial m_{k}}
\sum_{\mbox{\boldmath $\sigma$}_{k}}
p_{t}(\mbox{\boldmath $\sigma$}_{k})
\left\{
\frac{1}{N}\sum_{i}
\sigma_{i}(k)
\right\}
\delta (m_{k}-m_{k}(\mbox{\boldmath $\sigma$}_{k})) \nonumber \\
\mbox{} & - &   
\frac{\partial}
{\partial m_{k}}
\sum_{\mbox{\boldmath $\sigma$}_{k}}
p_{t}(\mbox{\boldmath $\sigma$}_{k})
\frac{1}{N}
\sum_{i}
\tanh [\beta \phi_{i}(\mbox{\boldmath $\sigma$}_{k} : \sigma_{i}(k+1))]
\delta (m_{k}-m_{k}(\mbox{\boldmath $\sigma$}_{k}))  \nonumber \\
\mbox{} & = & 
\frac{\partial}{\partial m_{k}}
\{
m_{k}P_{t}(m_{k})
\}  -  
\frac{\partial}{\partial m_{k}}
\left\{
P_{t}(m_{k})
\frac{1}{N}
\sum_{i}
\tanh [\beta \phi_{i}(\mbox{\boldmath $\sigma$}_{k} : \sigma_{i}(k+1))]
\right\} 
\label{eq:dPmkdt}
\end{eqnarray}
by neglecting ${\cal O}(N^{-1})$ term. 
To continue the derivation of deterministic flow equations of order parameters, 
we use the following assumption: 
\begin{eqnarray}
\lim_{N \to \infty} 
\frac{1}{N}
\sum_{i}
\tanh [\beta \phi_{i}(\mbox{\boldmath $\sigma$}_{k} :  \sigma_{i}(k+1))]
 & = & 
 \langle \tanh[\beta \phi (k)] \rangle_{\backslash \sigma (k)}
 \label{eq:tanh}
 \end{eqnarray}
 where we defined effective single site local field 
 $\beta \phi (k) \equiv  
 (\beta J/M )m_{k} + B \sigma (k+1)$ 
 and the average 
 \begin{eqnarray*}
 \langle \cdots \rangle_{\backslash \sigma (k)} & \equiv & 
 \lim_{M \to \infty}
 \sum_{\sigma (1)}
 \cdots \sum_{\sigma (k-1)}
 \sum_{\sigma (k+1)}
 \cdots \sum_{\sigma (M)}
 (\cdots) p(\sigma (1),\cdots,\sigma (k-1),\sigma (k+1),\cdots, \sigma (M)) \nonumber \\
 \mbox{} & = & 
  \lim_{M \to \infty}
  \frac{
 \sum_{\sigma (1)}
 \cdots \sum_{\sigma (k-1)}
 \sum_{\sigma (k+1)}
 \cdots \sum_{\sigma (M)}
 (\cdots)\,
 {\exp} [\beta \sum_{l \neq k}^{M} \phi (l)\sigma (l)]}
 {
 \sum_{\sigma (1)}
 \cdots \sum_{\sigma (k-1)}
 \sum_{\sigma (k+1)}
 \cdots \sum_{\sigma (M)}
 {\exp} [\beta \sum_{l \neq k}^{M} \phi (l)\sigma (l)]}. 
 \end{eqnarray*}
 Namely, here we shall assume that in the thermodynamical limit 
 $N \to \infty$, 
 the physical quantity in a particular choice (realization) of the Trotter slice, say,  
 the $k$-th Trotter slice 
 $N^{-1}\sum_{i}
\tanh [\beta \phi_{i}(\mbox{\boldmath $\sigma$}_{k} : \sigma_{i}(k+1))]$ 
is identical to its own average over all possible paths in the imaginary-time axis for 
both `past' $\{\sigma (1),\sigma (2),\cdots,\sigma (k-1)\}$ and 
`future' $\{\sigma (k+1),\cdots,\sigma (\infty)\}$. 
The weight of each path is proportional to 
\begin{eqnarray*}
{\exp}\left[
\beta \sum_{l \neq k}^{M} \phi (l)\sigma (l)
\right] & = & 
{\exp}
\left[
\frac{\beta J}{M}
\sum_{l \neq k}
m_{l}\sigma (l) + 
B \sum_{l \neq k} \sigma (l)\sigma (l+1)
\right]. 
\end{eqnarray*}
In order to grasp the physical meaning 
of the quantity (\ref{eq:tanh}), it might be helpful for us to 
notice that $\tanh [\beta \phi (k)]$ can be rewritten as 
\begin{eqnarray*}
\tanh [\beta \phi (k)] & = & 
\frac{\sum_{\sigma (k) =\pm 1} \sigma (k) {\exp}[\beta \phi (k)\sigma (k)]}
{
\sum_{\sigma (k)=\pm 1} 
{\exp}[\beta \phi (k)\sigma (k)]} = 
\sum_{\sigma (k)=\pm 1} \sigma (k) p(\sigma (k)|\sigma (1),\cdots,\sigma (k-1)). 
\end{eqnarray*}
Thus, the quantity $\tanh [\beta \phi (k)]$ means a conditional expectation over 
the effective single spin $\sigma (k)$ (in the thermodynamic limit $N \to \infty$) for 
a given `past' $\{\sigma (1),\cdots,\sigma (k-1)\}$ and 
it becomes a function of $\sigma (k+1)$. 
After averaging over the `past' with the weight 
$P(\sigma (1),\cdots, \sigma (k))$ and `future' with the weight 
$P(\sigma (k+1),\cdots,\sigma (\infty))$, we obtain 
\begin{eqnarray*}
\langle \tanh [\beta \phi (k)] \rangle_{\backslash \sigma (k)} & = & 
\lim_{M \to \infty} 
\frac{{\rm tr}_{\{\sigma\}}
\sigma (k) \exp [\beta \sum_{l=1}^{M} \phi (l)\sigma (l)]}
{{\rm tr}_{\{\sigma\}}
\exp [\beta \sum_{l=1}^{M} \phi (l)\sigma (l)]} \nonumber \\
\mbox{} & = & 
\lim_{M \to \infty} 
\frac{{\rm tr}_{\{\sigma\}}
\sigma (k) \exp [\frac{\beta J}{M}\sum_{l=1}^{M}m_{l}\sigma (l) + B\sum_{l=1}^{M}\sigma (l)\sigma (l+1)]}
{{\rm tr}_{\{\sigma\}}
\exp  [\frac{\beta J}{M}\sum_{l=1}^{M}m_{l}\sigma (l) + B\sum_{l=1}^{M}\sigma (l)\sigma (l+1)]}  
\equiv  \langle \sigma (k) \rangle_{path}
\end{eqnarray*}
where we defined the summation of all possible paths by ${\rm tr}_{\{\sigma\}} (\cdots) \equiv 
\sum_{\sigma (1)=\pm 1} \cdots \sum_{\sigma (M)=\pm 1} (\cdots)$. 
We should notice that 
in the classical limit $B \to \infty$ ($\Gamma \to 0$), 
the weight of the path in the imaginary-time axis is dominated by 
$\sigma (1) = \cdots = \sigma (M), m_{1}=\cdots m_{k}=m$ 
as we always see it  in the path integral approach of quantum mechanics in the limit of the zero Planck constant as $\hbar \to 0$ \cite{Feynman,Kleinert} , namely, 
$P(\sigma (1), \cdots, \sigma (M))=
\prod_{l=1}^{M}
\delta (\sigma (l)-\sigma)$ and we have 
\begin{eqnarray}
\lim_{B \to 0\, (\Gamma \to 0)}
\langle \sigma (k) \rangle_{path} & = & 
\langle \sigma (k) \rangle_{classical\, path} =  
\frac{\sum_{\sigma} \sigma\, {\exp}(\beta Jm \sigma)}
{\sum_{\sigma} {\exp}(\beta Jm \sigma)} = \tanh (\beta Jm). 
\end{eqnarray}
Substituting the result into (\ref{eq:dPmkdt}) and carrying out the integral 
with respect to $m_{k}$  
after multiplying the $m_{k}$, we immediately 
obtain the spontaneous magnetization flow for the dynamical Ising model; 
$dm/dt=-m + \tanh (\beta Jm)$. Near the critical point, it behaves as 
$dm/dt = -(1-\beta J)m -(\beta J m)^{3}/3 +{\cal O}(m^{5})$. 
From this equation, we easily find that spontaneous magnetization 
shows well-known time-dependent behaviour  
$m(t) \simeq m(0)\,{\rm e}^{-t/(1-\beta J)^{-1}}$ 
around the critical point $\beta \simeq \beta_{c} =J^{-1}$, 
and at the critical point, the relaxation time diverges as $(1-\beta_{c} J)^{-1}$ resulting in 
$m(t) \simeq t^{-1/2}$ (critical slowing down with dynamical exponent $\nu^{'}=1/2$). 

Therefore quantum fluctuation comes from 
the finite $B$ ($\Gamma >0$).  For the quantum case, the equation (\ref{eq:dPmkdt}) leads to  
\begin{eqnarray}
\frac{d P_{t}(m_{k})}
{dt} & = & 
\frac{\partial}{\partial m_{k}}\{m_{k}P_{t}(m_{k})\}
-\frac{\partial}{\partial m_{k}}\{
P_{t}(m_{k}) \langle \sigma (k) \rangle_{path}
\}
\label{eq:dPmkdt2}
\end{eqnarray}
In order to obtain the deterministic equation of order parameter, we should use the static 
approximation $m_{k} = m\,\,\,\, \forall (k)$.  
Under this assumption and using the inverse procedure of the Suzuki-Trotter 
decomposition (\ref{eq:ST}) : 
\begin{eqnarray*}
\lim_{M \to \infty} 
Z_{M} & \equiv &  
\lim_{M \to \infty}
{\rm tr}_{\{\sigma\}}
{\exp}
\left[
\frac{\beta Jm}{M}
\sum_{k}\sigma (k) + 
B \sum_{k}\sigma (k)\sigma (k+1)
\right] =  
{\rm tr}\,{\exp}[\beta Jm \sigma_{z} + \beta \Gamma \sigma_{x}]
\end{eqnarray*}
 we immediately have  $\langle \sigma (k) \rangle_{path}  =  
 \lim_{M \to \infty}
 \langle M^{-1} \sum_{k}\sigma (k) \rangle_{path}$ as 
\begin{eqnarray*}
 \langle \sigma (k) \rangle_{path} & = &  
 \lim_{M \to \infty} 
\frac{{\rm tr}_{\{\sigma\}}
\frac{1}{M}\sum_{k} \sigma (k) \exp [\frac{\beta Jm}{M}\sum_{l=1}^{M}\sigma (l) + B\sum_{l=1}^{M}\sigma (l)\sigma (l+1)]}
{{\rm tr}_{\{\sigma\}}
\exp  [\frac{\beta Jm}{M}\sum_{l=1}^{M}\sigma (l) + B\sum_{l=1}^{M}\sigma (l)\sigma (l+1)]}  \nonumber \\
\mbox{} & = & 
\lim_{M \to \infty} 
\frac{\partial \log Z_{M}}{\partial (\beta J m)} =   
\frac{Jm}{\sqrt{(Jm)^{2}+\Gamma^{2}}}
\tanh 
\beta 
\sqrt{(Jm)^{2}+\Gamma^{2}} 
\end{eqnarray*}
and equation (\ref{eq:dPmkdt2}) leads to  
\begin{eqnarray*}
\frac{P_{t}(m)}
{dt} & = & 
\frac{\partial}{\partial m}
\{
m P_{t}(m)
\}
-
\frac{\partial}{\partial m}
\left\{
P_{t}(m)
\frac{Jm}{\sqrt{(Jm)^{2}+\Gamma^{2}}}
\tanh 
\sqrt{(Jm)^{2}+\Gamma^{2}}
\right\} 
\label{eq:use0}. 
\end{eqnarray*}
Finally, substituting the form $P_{t}(m)=
\delta (m-m(t))$ and 
making the integral by part with respect to 
$m$ after multiplying itself $m$,  we obtain 
the following deterministic equation.  
\begin{eqnarray}
\frac{dm}{dt} & = & 
-m  + 
\frac{Jm}{\sqrt{(Jm)^{2}+\Gamma^{2}}}
\tanh \beta 
\sqrt{(Jm)^{2}+\Gamma^{2}} 
\label{eq:dyn}
\end{eqnarray}
It is easy to see that the steady state $dm/dt=0$ is nothing but 
the equilibrium state described by the equation of state 
$m  = 
Jm\{\sqrt{(Jm)^{2}+\Gamma^{2}}
\}^{-1}
\tanh \beta 
\sqrt{(Jm)^{2}+\Gamma^{2}}$. 
\begin{figure}[ht]
\begin{center}
\includegraphics[width=7.9cm]{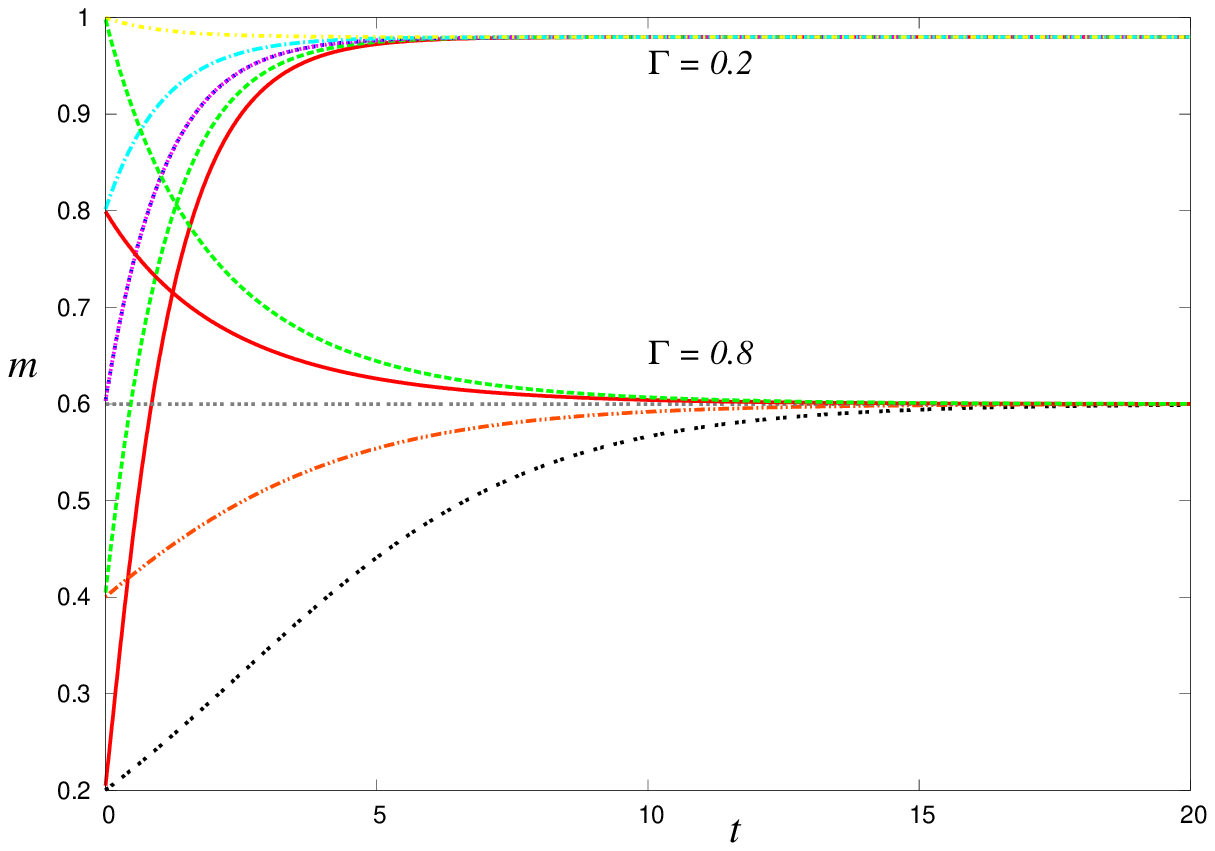}
\includegraphics[width=7.9cm]{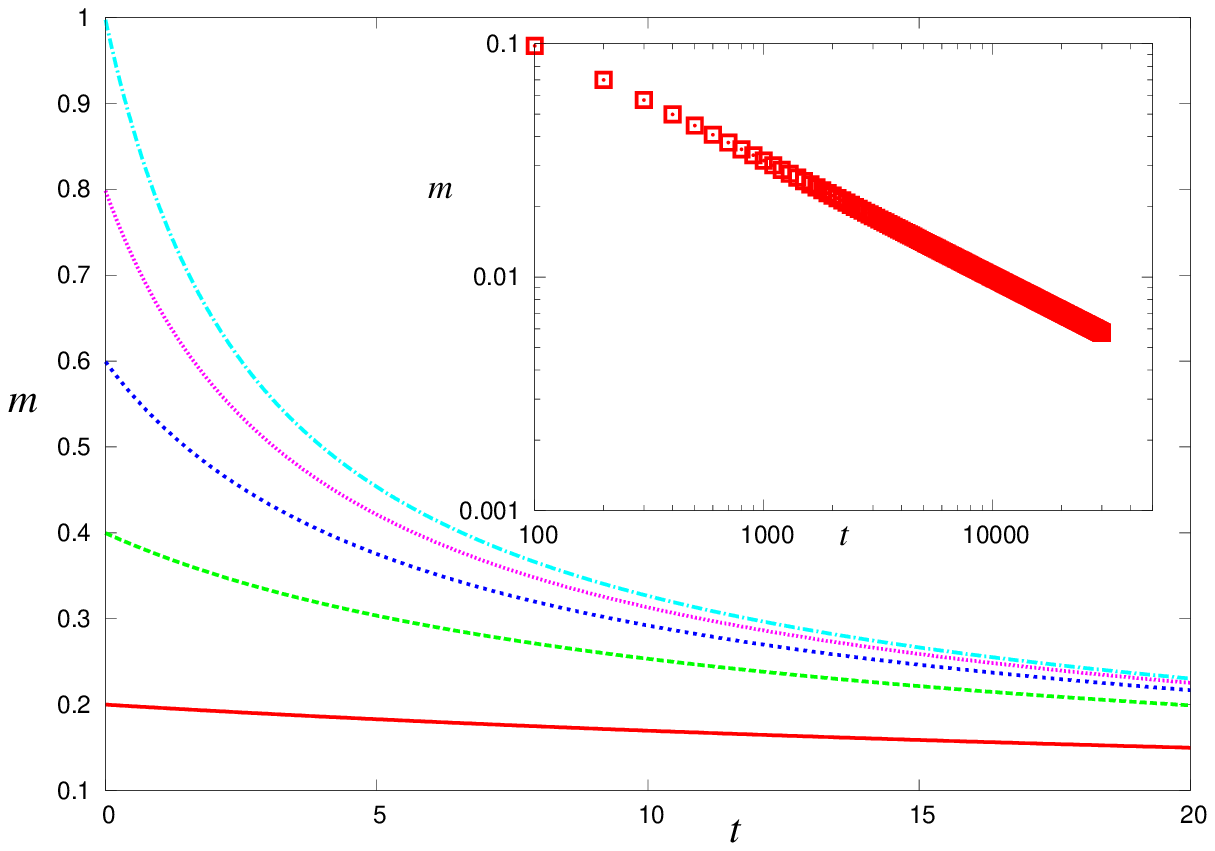}
\end{center}
\caption{\footnotesize 
Typical behaviour of zero-temperature 
dynamics described by (\ref{eq:dyn}) with $\beta =\infty$ 
far from the critical point $\Gamma_{c}=J=1$ of quantum phase transition (left). 
The right panel denotes 
the zero-temperature dynamics at the critical point. 
The inset shows the log-log plot of $m(t)$ indicating that 
the dynamical exponent in the critical slowing down is $\nu^{'}=1/2$.  
}
\label{fig:fg1}
\end{figure}
In Figure \ref{fig:fg1}(left), 
we plot the typical behaviour of zero-temperature 
dynamics (equation (\ref{eq:dyn}) with $\beta =\infty$) 
far from the critical point $\Gamma_{c}=J=1$ of quantum phase transition. 
 We easily find that the dynamics exponentially converges to the steady state. 
The right panel denotes 
the zero-temperature dynamics at the critical point. 
The inset shows the log-log plot of $m(t)$ indicating that 
the dynamical exponent in the critical slowing down is $\nu^{'} = 1/2$.  
This fact is directly confirmed from equation (\ref{eq:dyn}) with $\beta=\infty$ near the critical point  
$dm/dt \simeq -(1-J\Gamma^{-1})m-(Jm)^{3}/2\Gamma^{3} +
{\cal O}(m^{5})$, namely, 
$m(t)$ behaves around the critical point $\Gamma \simeq \Gamma_{c}=J$ 
as $m(t) = m(0)\,{\rm e}^{-t/(1-J\Gamma^{-1})^{-1}}$. 
At the critical point, the relaxation time 
diverges as  $\tau_{\Gamma} \equiv (1-J\Gamma_{c}^{-1})^{-1}$ resulting in the critical slowing 
down as $m (t) \simeq {\rm e}^{-t/\tau_{\Gamma}} \to m(t) \simeq t^{-\nu^{'}}$, $\nu^{'}=1/2$. 
Of course, the exponent is 
the same as that of the `mean-field model' universality class.   
\subsection{On the validity of static approximation}
Without the static approximation, the following deterministic flow 
equations for each Trotter slice is obtained by substituting the form $P_{t}(m_{k}) = 
\delta (m_{k}-m_{k}(t))$ and using the same way as deriving (\ref{eq:dyn}) as follows. 
\begin{eqnarray}
\frac{dm_{k}}{dt} & = & 
-m_{k} + 
\lim_{M \to \infty} 
\frac{{\rm tr}_{\{\sigma\}}
\sigma (k) \exp [\frac{\beta J}{M}\sum_{l=1}^{M}m_{l}\sigma (l) + B\sum_{l=1}^{M}\sigma (l)\sigma (l+1)]}
{{\rm tr}_{\{\sigma\}}
\exp  [\frac{\beta J}{M}\sum_{l=1}^{M}m_{l}\sigma (l) + B\sum_{l=1}^{M}\sigma (l)\sigma (l+1)]} 
\label{eq:dmkdt} 
\end{eqnarray}
Obviously, the equation (\ref{eq:dmkdt}) is symmetric for the choice of $k$ as long as we use the 
periodic boundary condition $\mbox{\boldmath $\sigma$}_{1}=\mbox{\boldmath $\sigma$}_{M+1}$. 
This might be a justification to assume that the static approximation is 
correct  at least for the present pure Ising system. 

\begin{figure}[ht]
\begin{center}
\includegraphics[width=7.9cm]{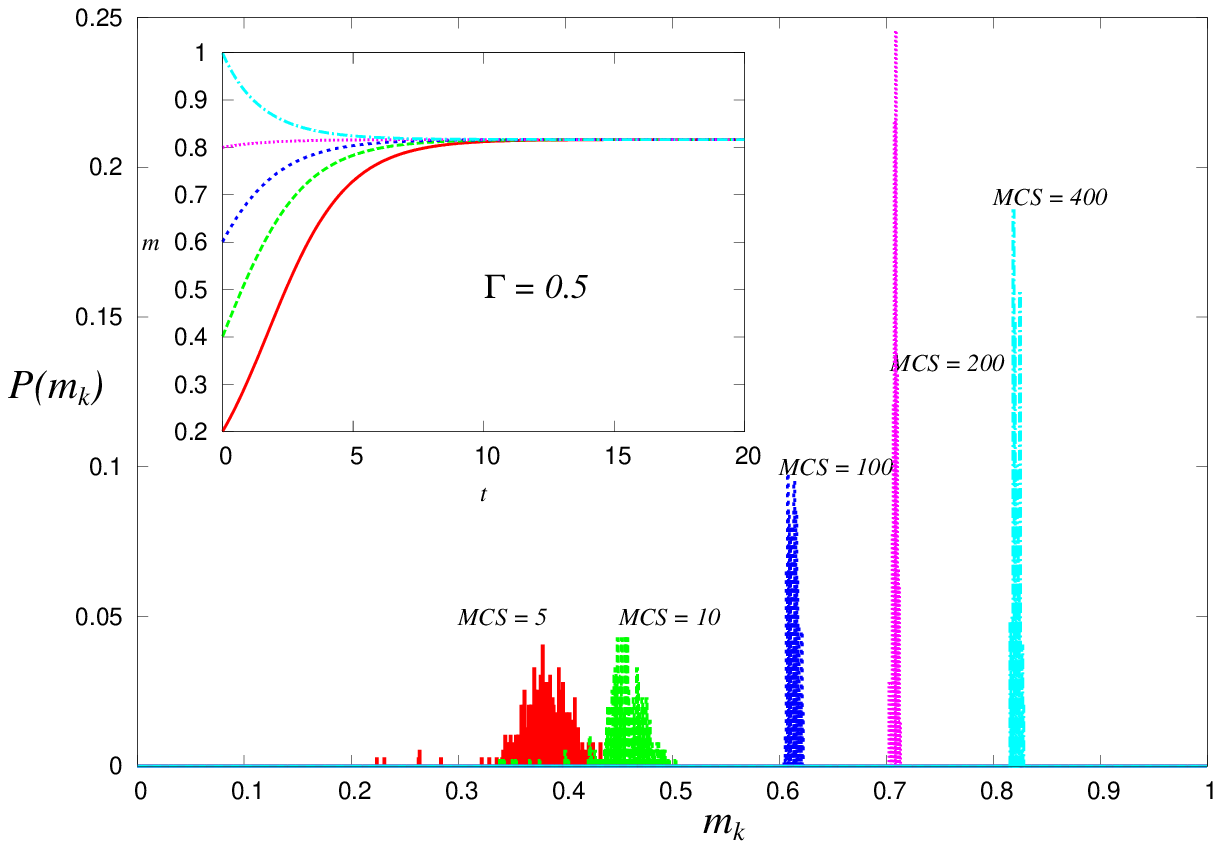}
\includegraphics[width=7.9cm]{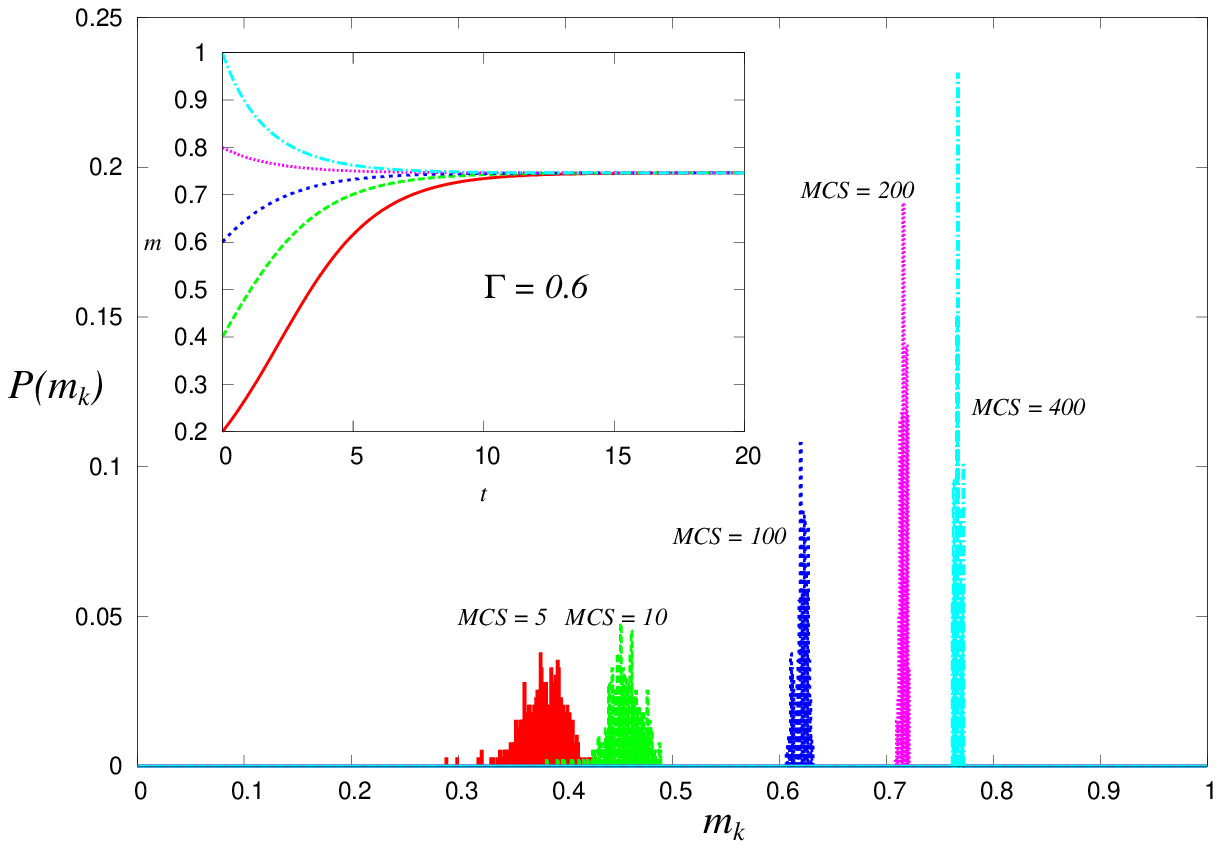}
\end{center}
\caption{\footnotesize 
Time evolution of the 
distribution $P(m_{k})$ calculated for 
finite size system with $N=M=400$. 
We choose the inverse temperature $\beta=2$ for 
$\Gamma=0.5$ (left) and $\Gamma=0.6$ (right). 
The inset in each panel 
denotes 
the deterministic flows of 
spontaneous magnetization 
calculated by (\ref{eq:dyn}) 
for corresponding parameter sets. 
}
\label{fig:fg2}
\end{figure}

To confirm this argument, we carry out computer simulation for finite size system 
having $N=400$ spins. We observe the 
time evolving process of the 
histogram $P(m_{k})$ which is calculated from the $M=N=400$ copies of 
the Trotter slices. We show the result in Figure \ref{fig:fg2}. 
In this simulation, we chose the initial configuration in each 
Trotter slice randomly (we set each spin variable $\sigma_{i}(k)$ to $+1$ 
with a fixed probability $p$) and 
choose the inverse temperature $\beta=2$ for 
$\Gamma=0.5$ and $\Gamma=0.6$. 
The time unit (the duration) of the update of $P(m_{k})$ is chosen as $1$ Monte Carlo step (MCS). 
From both panels in Figure \ref{fig:fg2}, 
we find that at the beginning, the $P(m_{k})$ is distributed due to the random set-up of the initial 
configuration, however, 
the fluctuation rapidly (eventually) shrinks leading up to 
the delta function around ${\rm MCS} \sim 100$. 
After that,  the $P(m_{k})$ evolves as a delta function with the peak 
located at the value of spontaneous magnetization 
which is explicitly indicated in the inset of each panel. 
It should be noted that 
we evaluated the value of order parameter 
at the time point in the Runge-Kutta method. 
Thus, the duration between the points to be evaluated is 
not the MCS but the Runge-Kutta step. 
Of course, some statistical errors for the finite system should be 
taken into account, however, the limited result here seems to support the validity of 
the static approximation even in the dynamical process.  
\section{Disordered systems: An application for Statistical-Mechanical Informatics}
It is easy for us to extend the above formulation to some class of disordered spin 
systems, that is to say, the infinite-range random field Ising model 
which is often used to check the performance of image restoration 
analytically \cite{NW,Inoue,Inoue2} as a bench mark test.
 
Here we consider a given original image $\mbox{\boldmath $\xi$} \equiv (\xi_{1},\cdots,\xi_{N})$ 
which is generated from the infinite-range ferromagnetic Ising model whose 
Gibbs measure (distribution for effective single spin) is described 
by $P(\xi_{i}) ={\rm e}^{\beta_{s} m_{0} \xi_{i}} /\{2 \cosh(\beta_{s}m_{0})\}$, where $m_{0}$ denotes spontaneous magnetization 
at temperature $\beta_{s}^{-1}$. 
A snapshot $\mbox{\boldmath $\xi$}$ from the distribution is degraded  by additive white Gaussian noise (AWGN)  
with mean $a_{0}\xi_{i}$ and variance $a^{2}$, namely, 
each pixel in the degraded image $\mbox{\boldmath $\tau$}=(\tau_{1},\cdots,\tau_{N})$ is obtained 
by $\tau_{i} = 
a_{0} \xi_{i} + a x_{i}$ 
with 
$x_{i} \sim {\cal N}(0,1)$. 
From the Bayesian inference view point, 
we assume that the posterior 
$P(\mbox{\boldmath $\sigma$}|\mbox{\boldmath $\tau$})$ 
(here we define $\mbox{\boldmath $\sigma$}$ as 
an estimate of the original image $\mbox{\boldmath $\xi$}$)  
might be proportional to 
the logarithm of the effective Hamiltonian 
(\ref{eq:Hamiltonian}) with $J_{ij} = J\,\,\forall(i,j)$ and $h \neq 0$. 
The first and the second terms appearing in the right hand side of 
(\ref{eq:Hamiltonian}) correspond to 
the prior distribution and 
the likelihood function, respectively. 
Whereas, the third term is introduced to utilize quantum fluctuation to 
construct the Bayes estimate for each pixel (`majority-vote decision' on each pixel), 
namely, 
${\rm sgn}(\langle \sigma_{i}^{z} \rangle)$. 

In this section, we 
attempt to describe the recovering process of original image 
through the deterministic 
flows of several relevant order-parameters and 
image restoration measure, namely, 
the overlap function 
${\cal M}\equiv N^{-1}\sum_{i}\xi_{i}\,{\rm sgn}(\langle \sigma_{i}^{z} \rangle)$. 

For the above set-up of the problem, the local field on the site $i$ in 
the $k$-th Trotter slice now leads to 
\begin{eqnarray*}
\phi_{i} (\mbox{\boldmath $\sigma$}_{k} : 
\sigma_{i}(k+1),\mbox{\boldmath $\tau$}) & = & 
\frac{J}{NM}\sum_{j}\sigma_{j}(k)+
\frac{h}{M} \tau_{i}\sigma_{i}(k) + 
B\sigma_{i}(k+1).
\end{eqnarray*}
As relevant order parameters,  we choose 
$\mu_{k}=N^{-1}\sum_{i}\tau_{i}\sigma_{i} (k)$ and 
magnetization 
$m_{k}$.  Then, we derive the differential equation with respect to 
$P_{t}(m_{k},\mu_{k}) = 
\sum_{\mbox{\boldmath $\sigma$}_{k}}
p_{t}(\mbox{\boldmath $\sigma$}_{k})\delta (m_{k}-
m_{k}(\mbox{\boldmath $\sigma$}_{k}))
\delta (\mu_{k}-
\mu_{k}(\mbox{\boldmath $\sigma$}_{k}))$ as follows.
\begin{eqnarray}
\frac{dP_{t}(m_{k},\mu_{k})}{dt} & = & 
\frac{\partial}{\partial m_{k}}
\{m_{k}P_{t}(m_{k},\mu_{k})\}
+
\frac{\partial}{\partial \mu_{k}}
\{
\mu_{k}P_{t}(m_{k},\mu_{k})\} \nonumber \\
\mbox{} & - & 
\frac{\partial}{\partial m_{k}}
\left\{
P_{t}(m_{k},\mu_{k})
\frac{1}{N}
\sum_{i}
\tanh 
[
\beta \phi_{i}(\mbox{\boldmath $\sigma$}_{k} : \sigma_{i}(k+1), 
\mbox{\boldmath $\tau$})] 
\right\} \nonumber \\
\mbox{} &  - & 
\frac{\partial}{\partial \mu_{k}}
\left\{
P_{t}(m_{k},\mu_{k})
\frac{1}{N}
\sum_{i}
\tau_{i} 
\tanh 
[
\beta \phi_{i}(\mbox{\boldmath $\sigma$}_{k} : \sigma_{i}(k+1), 
\mbox{\boldmath $\tau$})] 
\right\}
\end{eqnarray}
By assuming 
the self-averaging properties on the following physical 
quantities over both all possible paths in the imaginary-time axis  
and input data; original images and degrading processes 
(a particular realization of the quantity is identical to the average 
value and its deviation from the average eventually vanishes 
in the limit $N \to \infty$), we have 
\begin{eqnarray*}
N^{-1}
\sum_{i}
\tanh [
\beta \phi_{i}(\mbox{\boldmath $\sigma$}_{k} : \sigma_{i}(k+1), 
\mbox{\boldmath $\tau$})]   & = &   
[ \langle \sigma (k) \rangle_{*path}]_{data} \\
N^{-1}
\sum_{i}
\tau_{i} 
\tanh [
\beta \phi_{i}(\mbox{\boldmath $\sigma$}_{k} : \sigma_{i}(k+1), 
\mbox{\boldmath $\tau$})]  & = &   
[ \tau \langle \sigma (k) \rangle_{*path}]_{data} 
\end{eqnarray*}
where we defined the two different kinds of the averages by  
\begin{eqnarray}
\langle \cdots \rangle_{*path} & \equiv & 
\lim_{M \to \infty} 
\frac{{\rm tr}_{\{\sigma\}}
(\cdots)  \exp [\frac{\beta}{M}\sum_{l=1}^{M}(Jm_{l}+h \tau) \sigma (l) + B\sum_{l=1}^{M}\sigma (l)\sigma (l+1)]}
{{\rm tr}_{\{\sigma\}}
\exp  [\frac{\beta}{M}\sum_{l=1}^{M}(Jm_{l}+h \tau) \sigma (l) + B\sum_{l=1}^{M}\sigma (l)\sigma (l+1)]}  \nonumber \\
\mbox{} [\cdots]_{data} & \equiv & 
\frac{\sum_{\xi}{\rm e}^{\beta_{s}m_{0}\xi}}
{2\cosh (\beta_{s} m_{0})}
\int_{-\infty}^{\infty}
(\cdots) \,{\exp}
\left[
-\frac{(\tau-a_{0}\xi)^{2}}{2a^{2}}
\right] d\tau
\label{eq:dataAve}.  
\end{eqnarray}
Under the static approximation, we obtain 
\begin{eqnarray*}
\mbox{} & & 
\frac{dP_{t}(m, \mu)}{dt}  =  
\frac{\partial}{\partial m}
\{m P_{t}(m, \mu)\}
+
\frac{\partial}{\partial \mu}
\{
\mu P_{t}(m, \mu)\} \nonumber \\
\mbox{} & - & 
\frac{\partial}{\partial m}
\left\{
P_{t}(m,\mu)
\frac{\sum_{\xi}{\rm e}^{\beta_{s}m_{0}\xi}}
{2\cosh (\beta_{s} m_{0})}
\int_{-\infty}^{\infty} 
Dx 
\frac{ \varXi_{m}^{(a,a_{0})}(\xi,x) \tanh 
\beta 
\sqrt{
\{\varXi_{m}^{(a,a_{0})}(\xi,x)\}^{2}+\Gamma^{2}}}
{\sqrt{
\{\varXi_{m}^{(a,a_{0})}(\xi,x)\}^{2}+\Gamma^{2}}}
\right\} \nonumber \\
\mbox{} & - & 
\frac{\partial}{\partial \mu}
\left\{
P_{t}(m, \mu)
\frac{\sum_{\xi}{\rm e}^{\beta_{s}m_{0}\xi}}
{2\cosh (\beta_{s} m_{0})}
\int_{-\infty}^{\infty} 
Dx 
\frac{ (a_{0}\xi + ax) \varXi_{m}^{(a,a_{0})}(\xi,x) \tanh 
\beta 
\sqrt{
\{\varXi_{m}^{(a,a_{0})}(\xi,x)\}^{2}+\Gamma^{2}}}
{\sqrt{
\{\varXi_{m}^{(a,a_{0})}(\xi,x)\}^{2}+\Gamma^{2}}}
\right\}
\end{eqnarray*}
with $\varXi_{m}^{(a,a_{0})}(\xi,x) \equiv 
Jm + ha_{0}\xi + hax$ and 
$Dx \equiv dx\,{\rm e}^{-x^{2}/2}/\sqrt{2\pi}$. 
Using the same way as the pure Ising system discussed in the previous section, 
we finally obtain the deterministic flow equations of the 
order-parameters $m$ and $\mu$ as follows. 
\begin{eqnarray}
\frac{dm}{dt} & = & -m + 
\frac{\sum_{\xi}{\rm e}^{\beta_{s}m_{0}\xi}}
{2\cosh (\beta_{s} m_{0})}
\int_{-\infty}^{\infty} 
Dx 
\frac{ \varXi_{m}^{(a,a_{0})}(\xi,x) \tanh 
\beta 
\sqrt{
\{\varXi_{m}^{(a,a_{0})}(\xi,x)\}^{2}+\Gamma^{2}}}
{\sqrt{
\{\varXi_{m}^{(a,a_{0})}(\xi,x)\}^{2}+\Gamma^{2}}} 
\label{eq:dmdt} \\
\frac{d\mu}{dt} & = & -\mu + 
\frac{\sum_{\xi}{\rm e}^{\beta_{s}m_{0}\xi}}
{2\cosh (\beta_{s} m_{0})}
\int_{-\infty}^{\infty} 
Dx 
\frac{ (a_{0}\xi + ax) \varXi_{m}^{(a,a_{0})}(\xi,x) \tanh 
\beta 
\sqrt{
\{\varXi_{m}^{(a,a_{0})}(\xi,x)\}^{2}+\Gamma^{2}}}
{\sqrt{
\{\varXi_{m}^{(a,a_{0})}(\xi,x)\}^{2}+\Gamma^{2}}}
\label{eq:dmudt}.
\end{eqnarray}
For the solution of the above deterministic flows 
$(m,\mu)$  at time $t$,  
the overlap between the original image and degraded image is measured by
\begin{eqnarray}
{\cal M}(m, \mu) & = & 
\frac{\sum_{\xi}{\rm e}^{\beta_{s} m_{0} \xi}}
{2\cosh (\beta_{s}m_{0})}
\int_{-\infty}^{\infty}
Dx \, {\rm sgn}[\hat{m}+(a_{0}\xi + ax)\hat{\mu}] 
\label{eq:overlap2}
\end{eqnarray}
where $(\hat{m},\hat{\mu})$  is a 
solution of the following coupled equations
\begin{eqnarray}
m & = & 
 \frac{\sum_{\xi}{\rm e}^{\beta_{s} m_{0} \xi}}
{2\cosh (\beta_{s}m_{0})}
\int_{-\infty}^{\infty}
Dx \, \tanh 
[\hat{m}+(a_{0}\xi + ax)\hat{\mu}] 
\label{eq:hatm} \\
\mu & = & 
 \frac{\sum_{\xi}{\rm e}^{\beta_{s} m_{0} \xi}}
{2\cosh (\beta_{s}m_{0})}
\int_{-\infty}^{\infty}
Dx \, [\hat{m}+(a_{0}\xi + ax) \hat{\mu}]\, \tanh 
[\hat{m}+(a_{0}\xi + ax)\hat{\mu}]  
\label{eq:hatmu}
\end{eqnarray}
for a given point on the trajectory $(m,\mu)$ at time $t$. 
To obtain the overlap function (\ref{eq:overlap2}) and (\ref{eq:hatm})(\ref{eq:hatmu}), 
we used the concept of 
dynamical replica theory (the DRT) \cite{Coolen1994,Coolen1996}, 
namely, `equipartitioning' and `self-averaging' of the ${\cal M}$ during the evolution in time. 
As the derivation is a bit complicated, we shall 
show the detail in \ref{sec:AppA}. 

\begin{figure}[ht]
\begin{center}
\includegraphics[width=7.9cm]{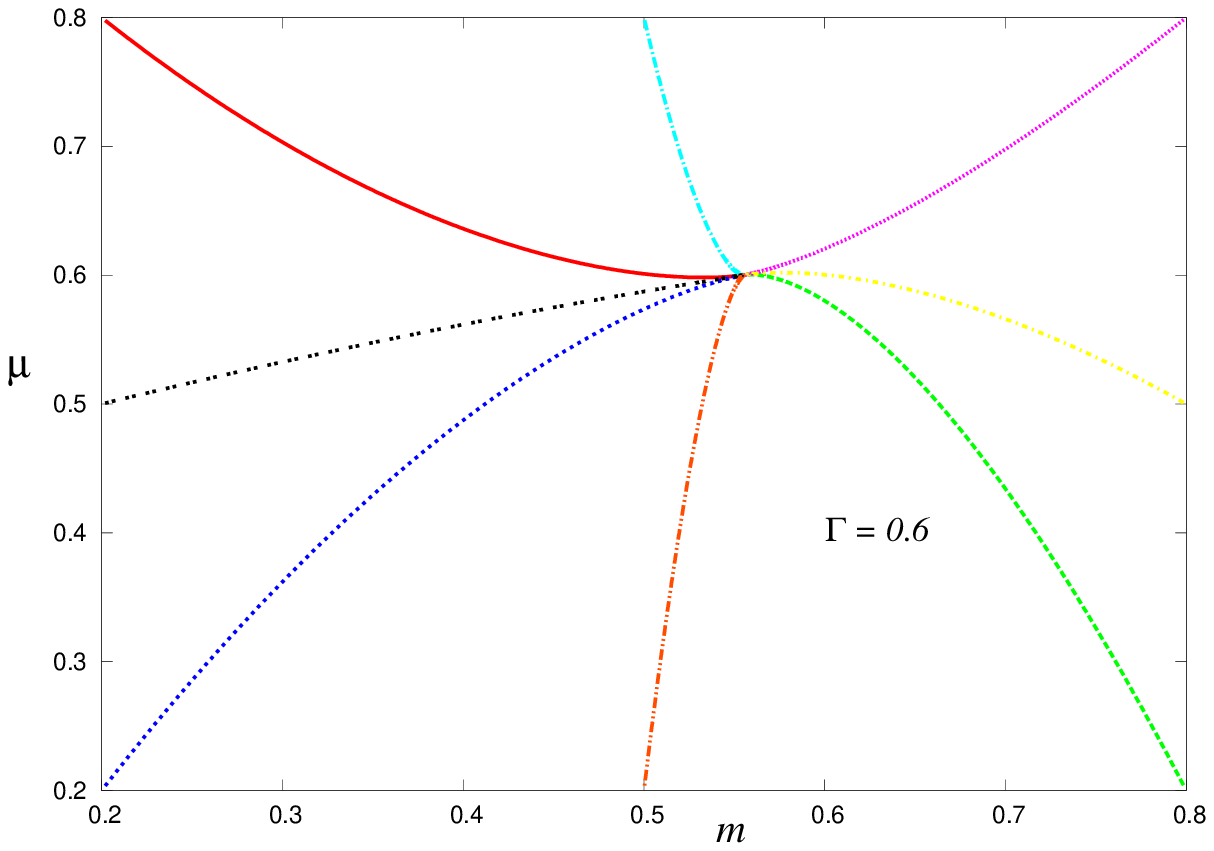}
\includegraphics[width=7.9cm]{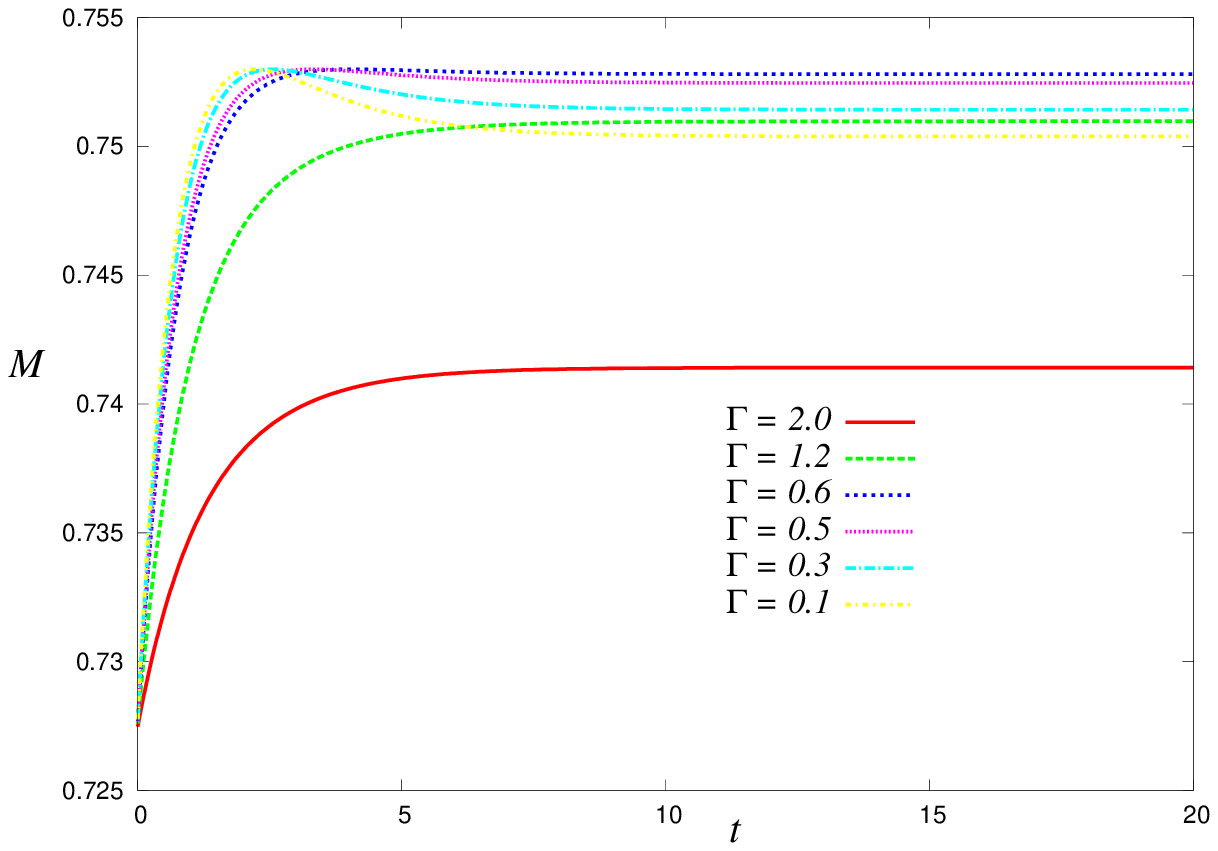}
\end{center}
\caption{\footnotesize 
Trajectories in the phase space $(m,\mu)$ are plotted for $\Gamma=0.6$ (left). 
The parameters are chosen as 
$\beta_{s}^{-1}=0.9$ ($m_{0}=0.523$), 
$a_{0}=a=1$ and the corresponding hyper-parameters as 
$h=a_{0}/a^{2}=1$ and $J^{-1}=\beta^{-1}$. 
The right panel shows 
the evolution of 
image restoration measure, 
overlap function 
${\cal M}$ in time for several values of 
the amplitude of quantum-mechanical tunneling $\Gamma$. }
\label{fig:fg3}
\end{figure}

We solve the equations  
(\ref{eq:dmdt})(\ref{eq:dmudt}) 
with 
(\ref{eq:overlap2})-(\ref{eq:hatmu}) 
numerically and show the results in 
Figure \ref{fig:fg3}.  
We choose the set of the parameters for  
the original image as 
$\beta_{s}^{-1}=0.9$ ($m_{0}=0.523$) and 
$a_{0}=a=1$ which means that  the corresponding 
optimal hyper-parameters are  
$h=a_{0}/a^{2}=1$ and $J^{-1}=\beta^{-1}$.  
We consider the zero-temperature 
restoration dynamics 
in which the fluctuation to make the 
Bayesian estimate is only quantum-mechanical one 
(it is controlled by the amplitude of quantum-mechanical tunneling $\Gamma$). 
In the left panel, the deterministic 
trajectories in the space $(m,\mu)$ are plotted for $\Gamma=0.6$. 
The state of system in which the image restoration is 
successfully achieved is in ferromagnetic phase. 
Thus, order-parameters $m$ and $\mu$ converge to 
the fixed point exponentially (there is no critical slowing down in this model system). 
In the right panel, we show the time evolution of 
the image restoration measure ${\cal M}$ 
for several values of $\Gamma$. 
From this panel, we find that some 
`non-monotonic' behaviour is observed at the initial stage of the 
dynamics when we fail to set the amplitude to its optimal value ($\Gamma \sim 0.6$). 
Similar behaviour was reported in the Bayesian 
image restoration via thermal (classical) fluctuation \cite{Ozeki}. 
\mbox{}

In the Bayesian framework, it is desired for us to obtain 
the estimate of each pixel and hyper-parameters simultaneously. 
In such case, we use the so-called EM algorithm based on the 
maximization of marginal likelihood criteria \cite{InoueTanaka}. 
The quantum-mechanical extension and the formulation 
presented here is applicable to the simultaneous estimation 
for both micro and macro parameters. 
\section{Concluding remarks}
In this paper, for a simplest quantum spin systems,  
we showed a formulation to 
describe the macroscopically 
deterministic flows of 
order parameters 
from the master equation whose transition probability is given by the Glauber-type. 
Under the static approximation,  differential equations with respect to macroscopic order 
parameters were explicitly obtained from the master equation describing 
the microscopic-law. 
In the steady state, we found that  the equations are 
identical to the saddle point equations for the equilibrium state of the same system. 
We also checked the validity of the static approximation by computer simulations 
and found that the result supports the validity of the approximation. 
 Several possible extensions of our approach to disordered spin systems 
for statistical-mechanical informatics was discussed. 
Especially, we used our procedure to evaluate the
decoding process of Bayesian image restoration. 
With the assistance of the concept of dynamical replica theory (the DRT), 
we derived the zero-temperature flow equation of image restoration measure 
showing some `non-monotonic'  behaviour in its time evolution.
Of course, by using the present approach, one can 
evaluate the `inhomogeneous'  Markovian stochastic process 
of quantum Monte Carlo method (in which amplitude $\Gamma$ is time-dependent) 
such as quantum annealing. 
In the next step of the present study, we are planning to extend this 
formulation to the probabilistic information processing described by spin glasses such as 
quantum Hopfield model \cite{Ma,Nishi1996} including a peculiar type of antiferromagnet \cite{Anjan}.
\appendix
\section{Derivation of overlap function:  A dynamical approach}
\label{sec:AppA}
In this appendix, we show the derivation 
of overlap function for image restoration problem (\ref{eq:overlap2})-(\ref{eq:hatmu}) 
by using a useful concept of dynamical replica theory (the DRT) \cite{Coolen1994,Coolen1996}. 

The problem here comes from the fact that 
the quantity ${\cal M}=N^{-1} \sum_{i} 
[\xi_{i} \,{\rm sgn} (\langle \sigma_{i} (k) 
\rangle_{path})]_{data}$ 
containing a single site expectation 
$\langle \sigma_{i} (k) 
\rangle_{path}$ 
is not observable in computer simulations of a single system. 
Therefore, we must introduce infinite number of `virtual copies' (`real replicas' or `ensembles' ) to evaluate 
the overlap function by using the law of large number. 
Namely, we evaluate the `real-time' dependence of the overlap function as follows. 
\begin{eqnarray*}
{\cal M}(t) & \equiv & 
\lim_{L, N \to \infty}
\frac{1}{N}
\sum_{i}
\xi_{i}\,
{\rm sgn}
\left(
\frac{1}{L}
\sum_{\rho =1}^{L}
\sigma_{it}^{\rho}(k)
\right) = 
\left[
\left\langle 
\xi\,
{\rm sgn}
\left(
\frac{1}{L}
\sum_{\rho =1}^{L}
\sigma_{1 t}^{\rho}(k)
\right)
\right\rangle_{copy} 
\right]_{data}  \nonumber \\
\mbox{} & = &  
{\cal M}(m_{k}(t),\mu_{k}(t))
\end{eqnarray*}
where $\rho=1,\cdots, L$ denote the copies 
of the original system in which 
$p_{t}(\mbox{\boldmath $\sigma$}^{\rho} (k))$ is the same. 
From the law of large number, 
we can expect $\lim_{L \to \infty} L^{-1} 
\sum_{\rho=1}^{L} \sigma_{it}^{l}(k)= 
\langle \sigma_{i}(k) \rangle_{copy}$. 
To calculate the expectation $\langle \cdots \rangle_{copy}$, we 
assume the equipartitioning in the $(m_{k},\mu_{k})$-subsells.  
Namely, 
explicit time-dependence of the expectation 
through $p_{t}(\mbox{\boldmath $\sigma$}^{\rho}(k))$ is now removed within the subsells as 
\begin{eqnarray*}
\langle \cdots \rangle_{copy} & = & 
\frac{\sum_{\{
\mbox{\boldmath $\sigma$}^{\rho}(k)\}}
\prod_{\rho=1}^{L}
p_{t}(\mbox{\boldmath $\sigma$}^{\rho}(k)) 
(\cdots)
\delta (m_{k}-m_{k}(\mbox{\boldmath $\sigma$}^{\rho}(k)))
\delta (\mu_{k}-\mu_{k}(\mbox{\boldmath $\sigma$}^{\rho}(k)))}
{\sum_{\{
\mbox{\boldmath $\sigma$}^{\rho}(k)\}}
\prod_{\rho=1}^{L}
p_{t}(\mbox{\boldmath $\sigma$}^{\rho}(k)) 
\delta (m_{k}-m_{k}(\mbox{\boldmath $\sigma$}^{\rho}(k))) 
\delta (\mu_{k}-\mu_{k}(\mbox{\boldmath $\sigma$}^{\rho}(k)))} \nonumber \\
\mbox{} & = & 
\frac{\sum_{\{
\mbox{\boldmath $\sigma$}^{\rho}(k)\}}
\prod_{\rho=1}^{L}
(\cdots)
\delta (m_{k}-m_{k}(\mbox{\boldmath $\sigma$}^{\rho}(k)))
\delta (\mu_{k}-\mu_{k}(\mbox{\boldmath $\sigma$}^{\rho}(k)))}
{\sum_{\{
\mbox{\boldmath $\sigma$}^{\rho}(k)\}}
\prod_{\rho=1}^{L} 
\delta (m_{k}-m_{k}(\mbox{\boldmath $\sigma$}^{\rho}(k))) 
\delta (\mu_{k}-\mu_{k}(\mbox{\boldmath $\sigma$}^{\rho}(k)))}. 
\end{eqnarray*}
Then, assuming the self-averaging on the ${\cal M}$,  we have 
${\cal M}_{L}(m_{k},\mu_{k})$ 
in the limit of $N \to \infty$ as 
\begin{eqnarray*}
\mbox{} &&
{\cal M}_{L} (m_{k},\mu_{k}) \nonumber \\
\mbox{} & = &
\lim_{N \to \infty}  
\left[
\frac{\sum_{\mbox{\boldmath $\sigma$}^{1}(k),\cdots, \mbox{\boldmath $\sigma$}^{L}(k)}
\prod_{\rho=1}^{L}
\delta (m_{k}-m_{k}(\mbox{\boldmath $\sigma$}^{\rho}(k)))
\delta (\mu_{k}-\mu_{k}(\mbox{\boldmath $\sigma$}^{\rho}(k)))
\xi_{1}
{\rm sgn}
\left(
\frac{1}{L}\sum_{\rho=1}^{L}
\sigma_{1}^{\rho}(k)
\right)}
{
\sum_{\mbox{\boldmath $\sigma$}^{1}(k),
\cdots, \mbox{\boldmath $\sigma$}^{L}(k)}
\prod_{\rho=1}^{L}
\delta (m_{k}-m_{k}(\mbox{\boldmath $\sigma$}^{\rho}(k)))
\delta (\mu_{k}-\mu_{k}(\mbox{\boldmath $\sigma$}^{\rho}(k)))
}
\right]_{data} \nonumber  \\
\mbox{} & = & 
\lim_{n\to 0}
\lim_{N \to \infty}
\left[
\sum_{\{\mbox{\boldmath $\sigma$}^{\rho \alpha}(k)\}}
\xi_{1}\,
{\rm sgn}
\left(
\frac{1}{L}\sum_{\rho=1}^{L}
\sigma_{1}^{\rho 1}(k)
\right)
\prod_{\alpha=1}^{n}
\prod_{\rho=1}^{L}
\delta (m_{k}-m_{k}(\mbox{\boldmath $\sigma$}^{\rho \alpha}(k)))
\delta (\mu_{k}-\mu_{k}(\mbox{\boldmath $\sigma$}^{\rho \alpha}(k)))
\right]_{data}
\end{eqnarray*}
where we used the fact 
\begin{eqnarray*}
\left[
\frac{{\rm tr}_{\{s\}}
{\cal P} \psi}
{{\rm tr}_{\{s\}}
{\cal P}}
\right]_{data} & = & 
\lim_{n \to 0}
\left[
\frac{({\rm tr}_{\{s\}}{\cal P}\psi)
({\rm tr}_{\{s\}}{\cal P})^{n}}
{{\rm tr}_{\{s\}}{\cal P}}
\right]_{data} = \lim_{n \to 0} \left[
{\rm tr}_{\{s^{\alpha}\}}\psi (s^{1}) \prod_{\alpha=1}^{n} {\cal P}(\{s^{\alpha}\})
\right]_{data}. 
\end{eqnarray*}
and introduced the replica index $\alpha=1,\cdots, n$ 
to carry out the average 
$[\cdots]_{data}$ by standard replica trick. 
It should be noted that 
$\mbox{\boldmath $\sigma$} \equiv 
(\sigma_{1}(k),\cdots,\sigma_{N}(k))$ and 
$\{ \cdots \}$ denotes the set of spin variables for all 
possible combinations of 
copies and replicas $(\rho,\alpha)$. 
By using the integral representation for 
the delta-function, we obtain 
\begin{eqnarray*}
{\cal M}_{L}(m_{k},\mu_{k}) & = & 
\lim_{n \to 0}
\lim_{N \to \infty}
\int_{-\infty}^{\infty}
\prod_{\rho \alpha}
\frac{d\hat{m}_{k}^{\rho\alpha}
d\hat{\mu}_{k}^{\rho \alpha}}
{2\pi N}
\,
{\rm e}^{-iN \sum_{\rho \alpha}
(
m_{k}\hat{m}_{k}^{\rho \alpha}
+
\mu_{k}\hat{\mu}_{k}^{\rho \alpha})
} \nonumber \\
\mbox{} & \times & 
\sum_{\{
\mbox{\boldmath $\sigma$}^{\rho\alpha}(k)\}}
\xi_{1} \,{\rm sgn}
\left(
\frac{1}{L}\sum_{\rho=1}^{L}
\sigma_{1}^{\rho 1}(k)
\right) 
{\rm e}^{-i \sum_{\rho\alpha}
\hat{m}_{k}^{\rho\alpha}
\sum_{i}\sigma_{i}^{\rho \alpha} (k)}
\left[
{\rm e}^{-i \sum_{\rho\alpha}
\hat{\mu}_{k}^{\rho\alpha}
\sum_{i}
\tau_{i} 
\sigma_{i}^{\rho \alpha} (k)}
\right]_{data}.
\end{eqnarray*}
Now, under the static approximation, 
the quantum fluctuation 
appears through the quantities $m_{k}=m$ and $\mu_{k}=\mu$ which 
obeys (\ref{eq:dmdt})(\ref{eq:dmudt}) at any time $t$. 
Therefore, from now on, we cancel the $k$-dependence. 
By simple transformation of the variables as 
$-i\hat{\mu}^{\rho\alpha} \mapsto 
\hat{\mu}^{\rho\alpha}, 
-i\hat{m}^{\rho\alpha} \mapsto 
\hat{m}^{\rho\alpha}$ and carrying out the 
data average $[\cdots]_{data}$ over the 
effective single site distribution (\ref{eq:dataAve}), one obtains 
\begin{eqnarray}
{\cal M}_{L}(m,\mu) & = & 
\lim_{n \to 0}
\lim_{N \to 0}
\int_{-\infty}^{\infty}
\prod_{\rho \alpha}
\frac{d\hat{m}^{\rho\alpha}
d\hat{\mu}^{\rho \alpha}}
{2\pi N}
\,
{\rm e}^{-N\sum_{\rho \alpha}
(m\hat{m}_{k}^{\rho \alpha}+
\mu \hat{\mu}_{k}^{\rho \alpha})
} \nonumber \\
\mbox{} & \times & \left\{
\frac{\sum_{\xi}{\rm e}^{\beta_{s}m_{0}\xi}}
{2\cosh(\beta_{s}m_{0})}
\int_{-\infty}^{\infty}
\prod_{\rho\alpha}
\left\{
2\cosh[\hat{m}^{\rho\alpha} + (a_{0}\xi +ax) \hat{\mu}^{\rho\alpha}]
\right\}
\right\}^{N} \nonumber \\
\mbox{} & \times & 
\frac{
\sum_{\sigma^{\rho\alpha}}
\xi_{1}{\rm sgn}
\left(
\frac{1}{L}
\sum_{\rho=1}^{L}
\sigma_{1}^{\rho 1}
\right) 
{\exp}[\sum_{\rho \alpha} \hat{m}^{\rho \alpha}\sigma^{\rho\alpha} + (a\xi+ax) \sum_{\rho\alpha} \hat{\mu}^{\rho\alpha} \sigma^{\rho \alpha}]}
{
\sum_{\sigma^{\rho\alpha}}
{\exp}[\sum_{\rho \alpha} \hat{m}^{\rho \alpha} + (a\xi+ax) \sum_{\rho\alpha} \hat{\mu}^{\rho\alpha} \sigma^{\rho \alpha}]} \nonumber \\
\mbox{} & \times & 
\lim_{n \to 0}
\lim_{N \to 0}
\int_{-\infty}^{\infty}
d\mbox{\boldmath $\hat{m}$}
d\mbox{\boldmath $\hat{\mu}$}
\,
{\exp}[N\Psi(\mbox{\boldmath $\hat{m}$},
\mbox{\boldmath $\hat{\mu}$})] \nonumber \\
\mbox{} & \times & 
\frac{
\sum_{\sigma^{\rho\alpha}}
\xi_{1}{\rm sgn}
\left(
\frac{1}{L}
\sum_{\rho=1}^{L}
\sigma_{1}^{\rho 1}
\right) 
{\exp}[\sum_{\rho \alpha} \hat{m}^{\rho \alpha} \sigma^{\rho\alpha} + (a\xi+ax) \sum_{\rho\alpha} \hat{\mu}^{\rho\alpha} \sigma^{\rho \alpha}]}
{
\sum_{\sigma^{\rho\alpha}}
{\exp}[\sum_{\rho \alpha} \hat{m}^{\rho \alpha} + (a\xi+ax) \sum_{\rho\alpha} \hat{\mu}^{\rho\alpha} \sigma^{\rho \alpha}]} 
\label{eq:ML}
\end{eqnarray}
with 
\begin{eqnarray*}
\Psi(\mbox{\boldmath $\hat{m}$},
\mbox{\boldmath $\hat{\mu}$}) &  \equiv & 
-\sum_{\rho\alpha}(m\hat{m}^{\rho\alpha} + \mu \hat{\mu}^{\rho\alpha}) + 
\log 
\left[
\frac{\sum_{\xi}{\rm e}^{\beta_{s}m_{0}\xi}}
{2\cosh(\beta_{s}m_{0})}
\int_{-\infty}^{\infty}
\prod_{\rho\alpha}
\left\{
2\cosh[\hat{m}^{\rho\alpha} + (a_{0}\xi +ax) \hat{\mu}^{\rho\alpha}]
\right\}
\right]. 
\end{eqnarray*}
We should notice that 
in the above expression of the overlap function ${\cal M}_{L}(m,\mu)$,  
the condition $\xi_{1}={\rm sgn}(L^{-1}\sum_{\rho=1}^{L}\sigma_{\rho 1}(k))$ 
leading to  `perfect image restoration', namely,  
$\lim_{L \to \infty}{\cal M}_{L}=1$ immediately gives 
$\lim_{n \to 0}
\lim_{N \to 0}
\int_{-\infty}^{\infty}
d\mbox{\boldmath $\hat{m}$}
d\mbox{\boldmath $\hat{\mu}$}
\,
{\exp}[N\Psi(\mbox{\boldmath $\hat{m}$},
\mbox{\boldmath $\hat{\mu}$})] =1$. 
Therefore, it is easy to find that the part 
$\lim_{n \to 0} \lim_{N \to 0}
\int_{-\infty}^{\infty}
d\mbox{\boldmath $\hat{m}$}
d\mbox{\boldmath $\hat{\mu}$}
\,
{\exp}[N\Psi(\mbox{\boldmath $\hat{m}$},
\mbox{\boldmath $\hat{\mu}$})]$  appearing in (\ref{eq:ML}) is a normalization factor. 
Thus, the overlap function derived by the above dynamical approach now leads to 
\begin{eqnarray*}
{\cal M}_{L}(m,\mu) & = &  
\frac{
\sum_{\sigma^{\rho\alpha}}
\xi_{1}\,{\rm sgn}
\left(
\frac{1}{L}
\sum_{\rho=1}^{L}
\sigma_{1}^{\rho 1}
\right) 
{\exp}[\sum_{\rho \alpha} \hat{m}^{\rho \alpha} \sigma^{\rho \alpha} + (a\xi+ax) \sum_{\rho\alpha} \hat{\mu}^{\rho\alpha} \sigma^{\rho \alpha}]}
{
\sum_{\sigma^{\rho\alpha}}
{\exp}[\sum_{\rho \alpha} \hat{m}^{\rho \alpha} \sigma^{\rho \alpha} + (a\xi+ax) \sum_{\rho\alpha} \hat{\mu}^{\rho\alpha} \sigma^{\rho \alpha}]} 
\end{eqnarray*}
where $\mbox{\boldmath $\hat{m}$}$ or $\mbox{\boldmath $\hat{\mu}$}$ should be chosen 
as a saddle point of the function $\Psi (\mbox{\boldmath $\hat{m}$},\mbox{\boldmath $\hat{\mu}$})$. 
Assuming the replica symmetric and the copy symmetric solution 
$\hat{m}^{\rho \alpha}=\hat{m},\hat{\mu}^{\rho\alpha}=\hat{\mu}\,\, \forall (\rho,\alpha)$, 
we obtain the function to be optimized at the saddle point. 
\begin{eqnarray*}
\Psi_{RS}=
\lim_{n \to 0}\lim_{N, L \to \infty}
\frac{\Psi (\mbox{\boldmath $\hat{m}$},\mbox{\boldmath $\hat{\mu}$})}{nL} = 
-m\hat{m}-\mu \hat{\mu}
+\frac{\sum_{\xi}{\rm e}^{\beta_{s}m_{0}\xi}}
{2\cosh(\beta_{s}m_{0})}
\int_{-\infty}^{\infty}Dx 
\cosh[\hat{m}+(a_{0}\xi + ax)\hat{\mu}]
\end{eqnarray*}
Obviously we find that 
the saddle point is obtained by the equations (\ref{eq:hatm}) and (\ref{eq:hatmu}). 
Then, the overlap function is evaluated as 
\begin{eqnarray*}
{\cal M}_{L}(m,\mu) & = & 
\frac{\sum_{\xi} \xi\, {\rm e}^{\beta_{s}m_{0}\xi}}
{2\cosh(\beta_{s}m_{0})}
\int_{-\infty}^{\infty}Dx 
\sum_{\sigma^{1},\cdots,\sigma^{L}}
{\rm sgn}
\left(
\frac{1}{L}
\sum_{\rho=1}^{L}
\sigma_{1}^{\rho 1}
\right)
\prod_{\rho=1}^{L}
\frac{{\rm e}^{\{\hat{m}+(a_{0}\xi + ax)\hat{\mu}\}\sigma_{1}^{\rho 1}}}
{2\cosh[\hat{m}+(a_{0}\xi + ax)\hat{\mu}]} \nonumber \\
\mbox{} & = & 
\int_{-\infty}^{\infty}
dz \,
\frac{\sum_{\xi} \xi\, {\rm e}^{\beta_{s}m_{0}\xi}}
{2\cosh(\beta_{s}m_{0})}
\int_{-\infty}^{\infty}Dx 
\sum_{\sigma^{1},\cdots,\sigma^{L}}
{\rm sgn}(z)
\delta 
\left(
z- \frac{1}{L}
\sum_{\rho=1}^{L}
\sigma_{1}^{\rho 1}
\right) \nonumber \\
\mbox{} & \times & 
\prod_{\rho=1}^{L}
\frac{{\rm e}^{\hat{m}+(a_{0}\xi + ax)\hat{\mu}}}
{2\cosh[\hat{m}+(a_{0}\xi + ax)\hat{\mu}]}  \equiv \int_{-\infty}^{\infty}dz 
\,{\rm sgn}(z){\cal P}_{L}(z)
\end{eqnarray*}
with 
\begin{eqnarray*}
{\cal P}_{L}(z) & \equiv & 
\frac{\sum_{\xi} \xi \, {\rm e}^{\beta_{s}m_{0}\xi}}
{2\cosh(\beta_{s}m_{0})}
\int_{-\infty}^{\infty}Dx 
\sum_{\sigma^{1},\cdots,\sigma^{L}}
\int_{-\infty}^{\infty}
\frac{dy}{2\pi}\,
{\rm e}^{iy(z-L^{-1}\sum_{\rho=1}^{L}\sigma_{1}^{\rho})}
\prod_{\rho=1}^{L}
\frac{{\rm e}^{\{\hat{m}+(a_{0}\xi + ax)\hat{\mu}\}\sigma_{1}^{\rho 1}}}
{2\cosh[\hat{m}+(a_{0}\xi + ax)\hat{\mu}]} \nonumber \\
\mbox{} & = & 
\int_{-\infty}^{\infty}
\frac{dy}{2\pi}{\rm e}^{iyz}
\frac{\sum_{\xi} \xi\, {\rm e}^{\beta_{s}m_{0}\xi}}
{2\cosh(\beta_{s}m_{0})}
\int_{-\infty}^{\infty}Dx 
\left\{
\frac{\cosh[\hat{m}+(a_{0}\xi + ax)\hat{\mu}-iL^{-1}y]}
{\cosh[\hat{m}+(a_{0}\xi +ax)\hat{\mu}]}
\right\}^{L} \nonumber \\
\mbox{} & = &  
\int_{-\infty}^{\infty}
\frac{dy}{2\pi}{\rm e}^{iyz}
\frac{\sum_{\xi} \xi\, {\rm e}^{\beta_{s}m_{0}\xi}}
{2\cosh(\beta_{s}m_{0})}
\int_{-\infty}^{\infty}Dx 
\left\{
\cos(yL^{-1})-i\sin(yL^{-1})
\tanh[\hat{m}+(a_{0}\xi + ax) \hat{\mu}]
\right\}^{L}.
\end{eqnarray*}
Thus, we have 
\begin{eqnarray*}
{\cal P} (z) & \equiv & 
\lim_{L \to \infty} {\cal P}_{L}(z)=  
\frac{\sum_{\xi} \xi\, {\rm e}^{\beta_{s}m_{0}\xi}}
{2\cosh(\beta_{s}m_{0})}
\int_{-\infty}^{\infty}Dx 
\int_{-\infty}^{\infty}
\frac{dy}{2\pi}\,
{\rm e}^{iy(z-\tanh[\hat{m}+(a_{0}\xi + ax) \hat{\mu}])} \nonumber \\
\mbox{} & = & 
\frac{\sum_{\xi} \xi\, {\rm e}^{\beta_{s}m_{0}\xi}}
{2\cosh(\beta_{s}m_{0})}
\int_{-\infty}^{\infty}Dx\,  
\delta (z-\tanh[\hat{m}+(a_{0}\xi + ax) \hat{\mu}]). 
\end{eqnarray*}
Substituting the result into 
${\cal M}_{L}$ and taking the limit of $L \to \infty$, 
we finally obtain
\begin{eqnarray*} 
{\cal M} (m,\mu) & = & 
\lim_{L \to \infty}{\cal M}_{L}(m,\mu) = 
\int_{-\infty}^{\infty}dz\,
{\rm sgn}(z) \lim_{L \to \infty} {\cal P}_{L}(z) = 
\int_{-\infty}^{\infty}dz\,
{\rm sgn}(z) {\cal P}(z) \nonumber \\
\mbox{} & = &    
\frac{\sum_{\xi} \xi\, {\rm e}^{\beta_{s}m_{0}\xi}}
{2\cosh(\beta_{s}m_{0})}
\int_{-\infty}^{\infty}Dx 
\,{\rm sgn} [\hat{m}+(a_{0}\xi + ax) \hat{\mu}]
\end{eqnarray*}
where we used the fact ${\rm sgn}(\tanh(x))={\rm sgn}(x)$. 
This result is nothing but equation (\ref{eq:overlap2}). 
As parameters $\hat{m}$ and $\hat{\mu}$ are 
related to the order-parameters $m$ and $\mu$ in 
equations (\ref{eq:hatm})(\ref{eq:hatmu}), 
overlap function ${\cal M}$ is influenced by quantum fluctuation 
through (\ref{eq:hatm})(\ref{eq:hatmu}) and 
the solution of equations (\ref{eq:dmdt})(\ref{eq:dmudt}). 
\ack
The present study was financially supported 
by {\it Grant-in-Aid 
Scientific Research on Priority Areas 
``Deepening and Expansion of Statistical Mechanical Informatics (DEX-SMI)" 
of The Ministry of Education, Culture, 
Sports, Science and Technology (MEXT)} 
No. 18079001 and 
{\it INSA (Indian National Science Academy) -  JSPS 
(Japan Society of Promotion of Science)  Bilateral Exchange Programme}. 
The author thanks Saha Institute of Nuclear Physics for their warm hospitality during his stay in India.  

\end{document}